\documentclass[twoside,leqno,twocolumn]{article}

\usepackage[letterpaper]{geometry}

\usepackage{ltexpprt}
\usepackage{hyperref}
\usepackage[most]{tcolorbox}
\usepackage{float}
\usepackage{booktabs}

\newtcolorbox{promptbox}{
    colback=gray!5,     
    colframe=black!75,  
    left=1em,          
    right=0.5em,       
    top=0.5em,         
    bottom=0.5em,      
    sharp corners,     
    boxrule=1pt         
}

\usepackage{amssymb}
\usepackage{graphicx}

\DeclareUnicodeCharacter{1D54F}{$\mathbb{X}$}

\begin{document}

\newcommand\relatedversion{}

\title{\Large 
User Migration across Multiple Social Media Platforms\relatedversion}
\newcommand{\authorspace}{\hspace{0.5em}} 
\renewcommand{\and}{\unskip\hspace{1em}} 
\author{Ujun Jeong\thanks{School of Computing and Augmented Intelligence, Arizona State University, \{ujeong1, anirmal1, kjha9, huanliu\}@asu.edu}\authorspace
\and Ayushi Nirmal\footnotemark[1]\authorspace
\and Kritshekhar Jha\footnotemark[1]\authorspace
\and Susan Xu Tang\thanks{Department of Economics, W. P. Carey School of Business, Arizona State University, Susan.Tang@asu.edu}\authorspace
\and H. Russell Bernard\thanks{Institute for Social Science Research, Arizona State University, asuruss@asu.edu}\authorspace
\and Huan Liu\footnotemark[1]}

\date{}

\maketitle


\fancyfoot[R]{\scriptsize{Copyright \textcopyright\ 2024 by SIAM\\
Unauthorized reproduction of this article is prohibited}}





\begin{abstract} \small\baselineskip=9pt After Twitter's ownership change and policy shifts, many users reconsidered their go-to social media outlets and platforms like Mastodon, Bluesky, and Threads became attractive alternatives in the battle for users. Based on the data from over 14,000 users who migrated to these platforms within the first eight weeks after the launch of Threads, our study examines: (1) distinguishing attributes of Twitter users who migrated, compared to non-migrants; (2) temporal migration patterns and associated challenges for sustainable migration faced by each platform; and (3) how these new platforms are perceived in relation to Twitter. Our research proceeds in three stages. First, we examine migration from a broad perspective, not just one-to-one migration. Second, we leverage behavioral analysis to pinpoint the distinct migration pattern of each platform. Last, we employ a Large Language Model (LLM) to discern stances towards each platform and correlate them with the platform usage. This in-depth analysis illuminates migration patterns amid competition across social media platforms.\end{abstract}

\vspace{0.25cm}




\section{Introduction}
In the years since the 1997 launch of Bolt and Six Degrees,  social media have become online hubs, offering many avenues for communication, entertainment, and information~\cite{auxier2021social}. Users are increasingly mobile, migrating between platforms as their needs, preferences, and interests evolve, driving intense competition among social media platforms for user attention. One example is the substantial migration from Twitter to Mastodon following Twitter's ownership change~\cite{jeong2023exploring, he2023flocking, cava2023drivers}. With the emergence of other platforms, like Threads and Bluesky, users are questioning whether their current ``cyber hometown'' is the best choice~\cite{valero2023thousands}.

Prior research has examined the motivations behind platform migration and typical behaviors of migrants--the pushes and pulls, as they are known in the social science literature~\cite{fiesler2020moving, hou2020understanding, jeong2023exploring}. Here, we extend this research to examine: (1) the varying engagement levels of migrating users based on their new platform choice; (2) the competitive dynamics between platforms seeking user attention and what influences their success; and (3) the perspectives of migrants towards each platform and how these perspectives associate with user behaviors.


\begingroup
\setlength{\abovecaptionskip}{0pt}
\setlength{\belowcaptionskip}{0pt}
\begin{figure}
\centering
\includegraphics[width=0.485\textwidth]{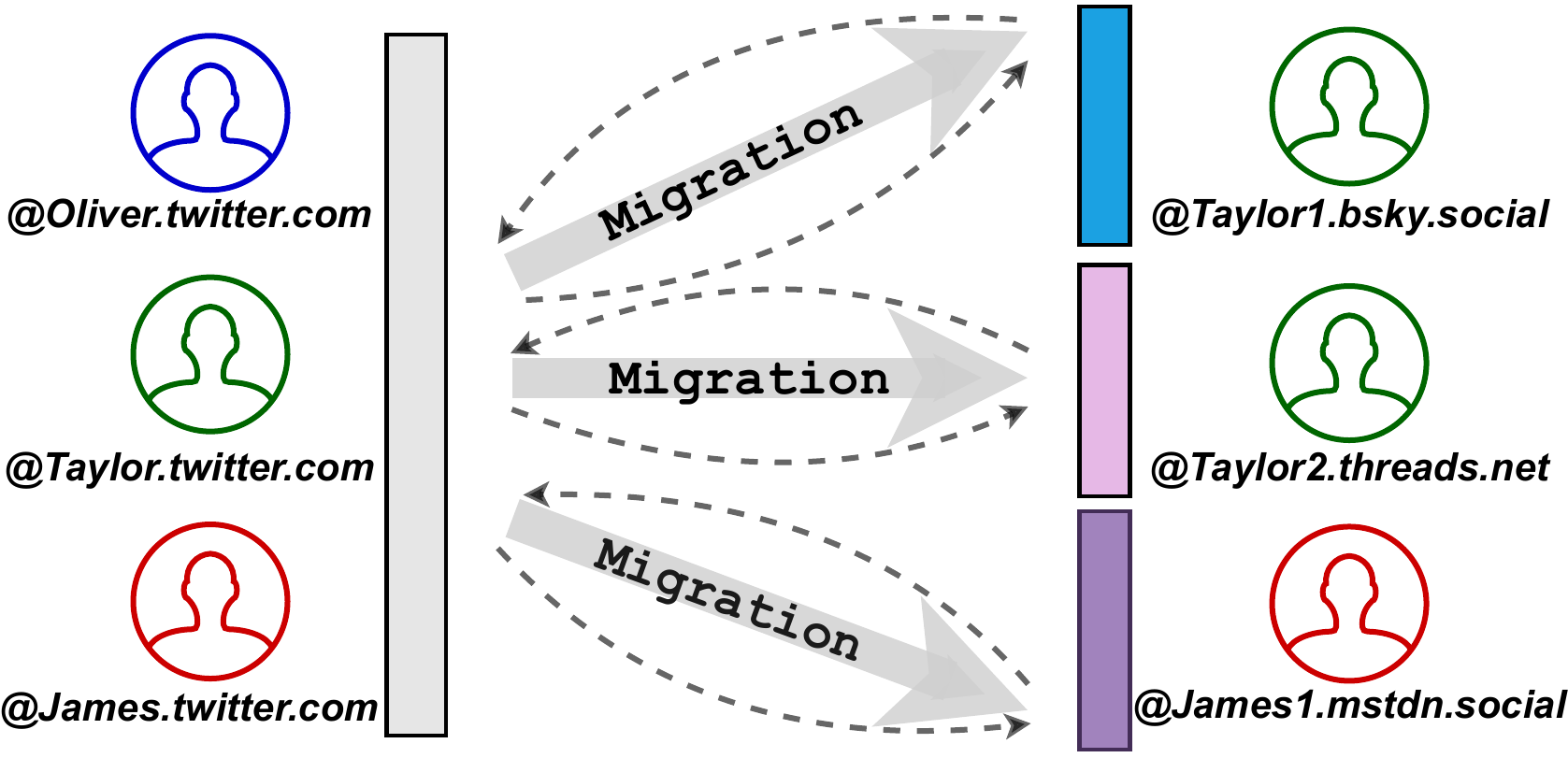}
\caption{\small The migration flow between Twitter and its alternatives: Mastodon, Bluesky, and Threads. The dashed lines represent the shift of user attention across these platforms.
}
\label {MgirationFlow}
\end{figure}
\endgroup


To collect data on platform migrants, we identified the account handles of 14,270 users who initiated migration from Twitter to Bluesky, Threads, and Mastodon, focusing on user profiles and their activities within the first eight weeks after the official launch of Threads on July 5, 2023. For those who did not migrate from Twitter, our sampling techniques leveraged network traffic analysis between Twitter and its counterparts, ensuring the chances of precise selection of non-migrants.


Our study is motivated by three questions:
\begin{itemize}
    \item \textbf{RQ1:} What characteristics distinguish migrant groups and non-migrants on Twitter?
    \item \textbf{RQ2:} What patterns of migration reveal the relationships between Twitter and other platforms?
    \item \textbf{RQ3:} After attempting to leave Twitter, did users sustain their engagement with their new platforms?
\end{itemize}

With respect to \textbf{RQ1}, we analyzed the behavioral traits of migrants to Bluesky, Threads, and Mastodon. We quantified their influence scores and compared them to those of non-migrants to determine the level of presence of these migrant groups on Twitter. This revealed that many users attempted to migrate despite their high level of presence on the prior platform.

Regarding \textbf{RQ2}, we examined the evolving migration patterns with users’ active status between Twitter and various pairs of platforms. Specifically, we measured the association between Twitter and other alternatives based on users' active status to understand the perceived relationships between the platforms by users.

Concerning \textbf{RQ3}, we assessed the perspectives of migrants on brand loyalty by analyzing the texts in their posts. Paradoxically, although the use of Twitter has increased over time, there was a distinct lack of loyalty expressed towards Twitter. We also examined the relationship between user loyalty and their patterns of platform selection to gain insights into the prevailing tendency for choosing a primary social media platform.

Our main contributions are as follows:
\begin{itemize}
\item We curated a dataset of 14,000+ users migrating from Twitter to Bluesky, Threads, and Mastodon, following those platforms' terms of service.
\item To our knowledge, this is the first study to examine the differences among multiple migrant groups (based on their chosen platforms) and contrast them with non-migrants.
\item Our comparative analysis of migration shows that, despite the rhetoric to the contrary, migrants have a strong inertia for Twitter over other platforms.
\end{itemize}



\section{Related Work}

\subsection{Human Migration and Platform Migration.}
Across the social sciences, the push-pull theory is widely used to explain human migration. The theory assumes that for every migration event, there are factors pushing people away from their home territory and factors pulling them towards a new home~\cite{levitt2007transnational, lee1966theory}. This can be applied to the migration of users between social media platforms~\cite{zengyan2009cyber}. Unlike physical movement, where one is constrained to a single location at a time, the digital world allows users to engage with multiple platforms simultaneously. Such a dynamic calls for a refined classification of online migrants~\cite{kumar2011understanding}. In economics, the concept of ``service switching'' parallels this phenomenon, portraying online users as shoppers exploring various platforms to find their preferred choice~\cite{givon1984variety, hou2011migrating}. Hence, the study of platform migration involves understanding the various factors for selecting social media platforms and the range of competing options in the market~\cite{azose2019estimation, fiesler2020moving}.

\subsection{Large-scale Online Migrations.}
Historically, substantial migrations between social media platforms have been observed, such as the shift from MySpace to Facebook~\cite{torkjazi2009hot} and from Facebook to Instagram~\cite{hou2020understanding}. Often, these shifts stem from perceived deficiencies in one platform and the emergence of superior features in another that better cater to user needs. The main drivers for this migration include push factors, such as low quality of service and bad experiences in social interactions, and pull factors, such as the presence of attractive 
new features and highly influential users on another platform.~\cite{hou2020understanding}. On Reddit, user migrations occurred as a response to moderation, such as deplatforming~\cite{monti2023online, rogers2020deplatforming} or policy changes~\cite{matias2016going}. Recently, Twitter's ownership change spurred a mass migration of users to Mastodon~\cite{cava2023drivers}. However, doubts arose about whether Mastodon could retain these users~\cite{jeong2023exploring}, prompting users to explore alternatives such as Bluesky and Threads.

Our research differs from previous studies on platform migration by examining the dynamics of user migration across multiple platforms, focusing on users' perspectives on inter-platform relationships. This study underlines potential factors for users reverting to their prior platform and challenges in platform migration.

\begin{figure}
\centering
\includegraphics[width=0.38\textwidth]{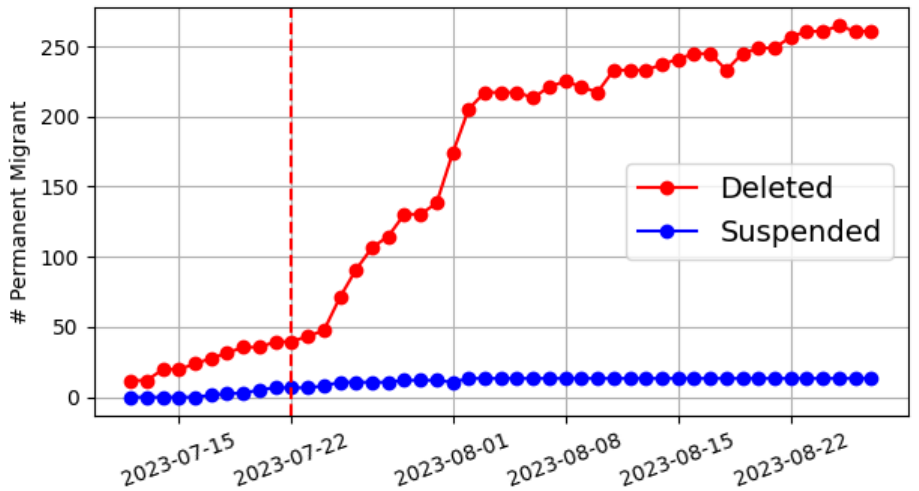}
\caption{\small The trend of deleted and suspended migrants' accounts on Twitter over time. The red dashed line marks the date Twitter announced its rebranding to ``$\mathbb{X}$''.} 
\label{permanent_migrants_trend}
\end{figure}

\section{Preliminaries}
In social media and migration studies, two types of migration are defined~\cite{hou2020understanding, levitt2007transnational}: (1) \textit{Permanent migration}, where users transition to a new platform, deactivate their original account, and exclusively engage on the new platform; and (2) \textit{Temporary migration}, where users maintain a presence on both platforms but switch their focus between them.

\vspace{0.1cm}
\paragraph{\textbf{Permanent Migration}} If user \( u \) was a member of platform \( p_1 \) at time \( t \) and is no longer on \( p_1 \) at time \( t' \), but has joined \( p_2 \), then user $u$ is considered to have permanently migrated from platform \( p_1 \) to \( p_2 \).

\vspace{0.1cm}
\paragraph{\textbf{Temporary Migration}} If user \( u \) is a member of \( p_1 \) before time \( t \) and is found on platforms \( p_1 \) and \( p_2 \) at a later time \( t' \). That user is considered to have temporarily migrated from platform \( p_1 \) to \( p_2 \).

Every day at 6 AM, we checked the status of 14,270 user profiles for signs of \textit{permanent migration} through actions like profile deletion or suspension on Twitter. As shown in Figure~\ref{permanent_migrants_trend}, only 1.8\% (270 out of 14,270) deleted their accounts on August 27, 2023. Most of these migrations occurred after Twitter's rebranding to ``$\mathbb{X}$'', implying that rebranding either precipitated or intensified this move~\cite{keller1993conceptualizing}.

In accordance with the General Data Protection Regulation (GDPR)\footnote{https://gdpr.twitter.com/en.html}, we refrained from gathering information on individuals who deactivated their Twitter accounts. Here, we only verified the existence of their accounts using the Twitter API. As a result, our study focuses on \textit{temporary migration}—users who migrated to new platforms but might return to Twitter later on.

\begin{table}
\begin{center}

\caption{\small Statistics on user migration from Twitter to other platforms. Bluesky and Threads operate on a single server or a cluster of servers, given their early stage of deployment.}
\label{migrants_statistics}

\resizebox{0.5\textwidth}{!}{
\begin{tabular}{ccc}
\toprule
 \textbf{Destination Platform} & \textbf{\# Migrant} & \textbf{\# Server (Domain)}  \\
\midrule
Bluesky & 1,062 & 1 (bsky.social) \\
\midrule
Threads & 679  & 1 (threads.net) \\
\midrule
Mastodon & 12,529 & 1,195 (mastodon.social, etc.)\\
\bottomrule
\end{tabular}
}
\end{center}
\end{table}

\section{Data Collection}

From July 1 to August 27, 2023, we identified a total of 14,270 migrants. After removing 270 migrants with deleted or suspended accounts, we were left with 14,000 migrants with unique handles. We further verified that none of these users had accounts on the destination platform before establishing their Twitter accounts.


\subsection{Collecting Migrants from Twitter to Destination Platforms.}



To accurately map Twitter users with their corresponding accounts on other social media platforms, we employed a platform-specific approach: (1) For Bluesky, by targeting keywords ``bsky.social'' and ``bsky.app'', we extracted relevant Bluesky handles from Twitter profiles; (2) For Threads, we used ``threads.net'' as our primary keyword filter, from which we derived associated Threads handles; and (3) For Mastodon, we began by gathering a complete list of 18,605 Mastodon server domains via the API from \textit{instances.social}. By using these domains as keywords, we identified Twitter profiles linked to Mastodon handles.

Noticing that Twitter users often include other account handles in their profiles, we examined their display names to pinpoint handles from different platforms. This approach avoids confusion, as handles mentioned in tweets may refer to other users~\cite{he2023flocking, jeong2023exploring}. Table~\ref{migrants_statistics} displays the count of detected migrants for each platform. The fewest migrations were noted from Twitter to Threads, likely the result of Twitter's recent action of hiding tweets containing URLs that link to Threads\footnote{https://www.washingtonpost.com/technology/2023/08/15/twitter-x-links-delayed/}.




\subsection{Collecting Non-migrants from Twitter.}

\begin{figure}
\centering
\includegraphics[width=0.5\textwidth]{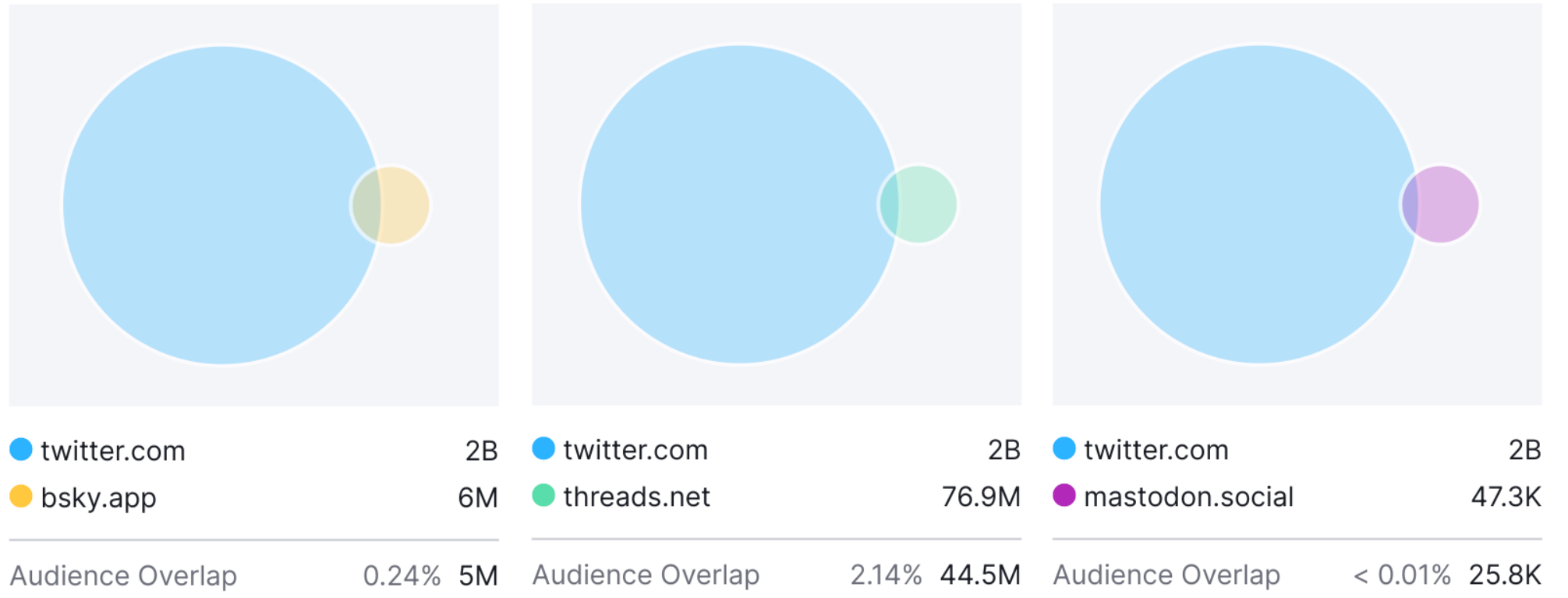}
\caption{\small Network traffic analysis for July-August 2023 between Twitter and the targeted domains. Overlaps show users accessing both domains estimated by Semrush.} 
\label{Audience_Overlap}
\end{figure}

We utilized Semrush\footnote{https://www.semrush.com/}, a tool designed for network traffic and competitor analysis to estimate non-migrants. The results are shown in Figure~\ref{Audience_Overlap}, which indicates users active across Twitter, along with Bluesky, Threads, and Mastodon. The maximum contribution from the targeted platforms to Twitter's total traffic is 2.48\%, when there are no overlaps among them. Based on this, we randomly sampled 20,000 unique active users who tweeted at least once on Twitter between July and August 2023 to minimize the inclusion of those migrant users to Bluesky, Threads, and Mastodon.


\subsection{Collecting Profiles and Posts of Migrants from Multiple Platforms.}

Within our study's timeframe, we gathered profiles and posts from users who migrated from Twitter to other platforms. For the migrants moving to Bluesky, we collected 98K posts from Twitter and 285K from Bluesky. For migrants moving to Threads, we collected 67K posts from Twitter and 10K posts from Threads. For migrants moving to Mastodon, we collected 190K posts from Twitter and 229K posts from Mastodon. The collected data were anonymized and also securely stored within a MongoDB database, protected by field-level encryption.






\begin{table}
 \begin{center}
  \caption{\small Ranking of migrant groups and non-migrants across five user metrics. The inequality symbol denotes a significant disparity, while the equality symbol indicates no significant disparity as assessed by ANOVA test at $p<0.05$.}
  \label{ANOVA_ranking}
  \footnotesize
  
\resizebox{0.5\textwidth}{!}{
  \begin{tabular}{cc}
   \toprule
   \textbf{User Metric} & \textbf{Statistical Disparities and Rankings of Means} \\ 
   \midrule
   \textit{followers\_count} & Threads $>$ Bluesky $>$ (Mastodon $=$ Non-migrant) \\
   \midrule
   \textit{friends\_count} & Threads $>$ Bluesky $>$ Non-migrant $>$ Mastodon \\
   \midrule
   \textit{listed\_count} & Bluesky $>$ Threads $>$ Mastodon $>$ Non-migrants \\
   \midrule
   \textit{favorites\_count} & (Non-migrant $=$ Bluesky) $>$ Threads $>$ Mastodon \\
   \midrule
   \textit{statuses\_count} & Non-migrant $>$ (Bluesky $=$ Threads) $>$ Mastodon \\
   \bottomrule
  \end{tabular}
  }
 \end{center}
\end{table}

We used the platforms' APIs. Twitter's official API provides comprehensive access to user profiles, tweets, retweets, and various metadata elements, ensuring a detailed view of user activities. Bluesky, aiming to develop a decentralized standard for social media, also offers its official API\footnote{https://atproto.com/guides/overview/} based on the AT protocol, which was instrumental in accessing public posts and profile details. Mastodon, being an open-source and federated platform, offers its official API\footnote{https://docs.joinmastodon.org/api/} based on the ActivityPub protocol. Since Threads does not currently have an official API, we manually collected the text contents of public user profiles and posts through the platform's web interface\footnote{https://www.threads.net/}, which was released on August 24, 2023.

\section{Distinguishing between Migrant Groups and Non-migrants on Twitter (RQ1)}

For this question, we analyzed a range of user characteristics, from basic metrics such as the number of followers to intricate measures of user influence. We then compared migrant groups and non-migrants on Twitter.

\subsection{Comparing Profile Metrics.}
We examined all the numerical metrics provided in Twitter's user profile object\footnote{https://developer.twitter.com/en/docs/twitter-api/data-dictionary/object-model/user}. The user profile metrics include \textit{followers\_count}, \textit{friends\_count}, \textit{listed\_count}, \textit{favorites\_count}, and \textit{statuses\_count}. We compared the variations in these metrics between migrants on different platforms (Bluesky, Threads, and Mastodon) and non-migrants on Twitter, using a multiple comparison analysis with ANOVA and Tukey's HSD post-hoc test. The table displays the statistical disparities and rankings of means, based on the platforms targeted by the migrants. In particular, the migrant groups of Threads and Bluesky consistently demonstrate stronger engagement in social connections compared to other groups, as indicated by higher \textit{followers\_count}, \textit{friends\_count}, and \textit{listed\_count}.


\begin{figure}
\centering
\includegraphics[width=0.47\textwidth]{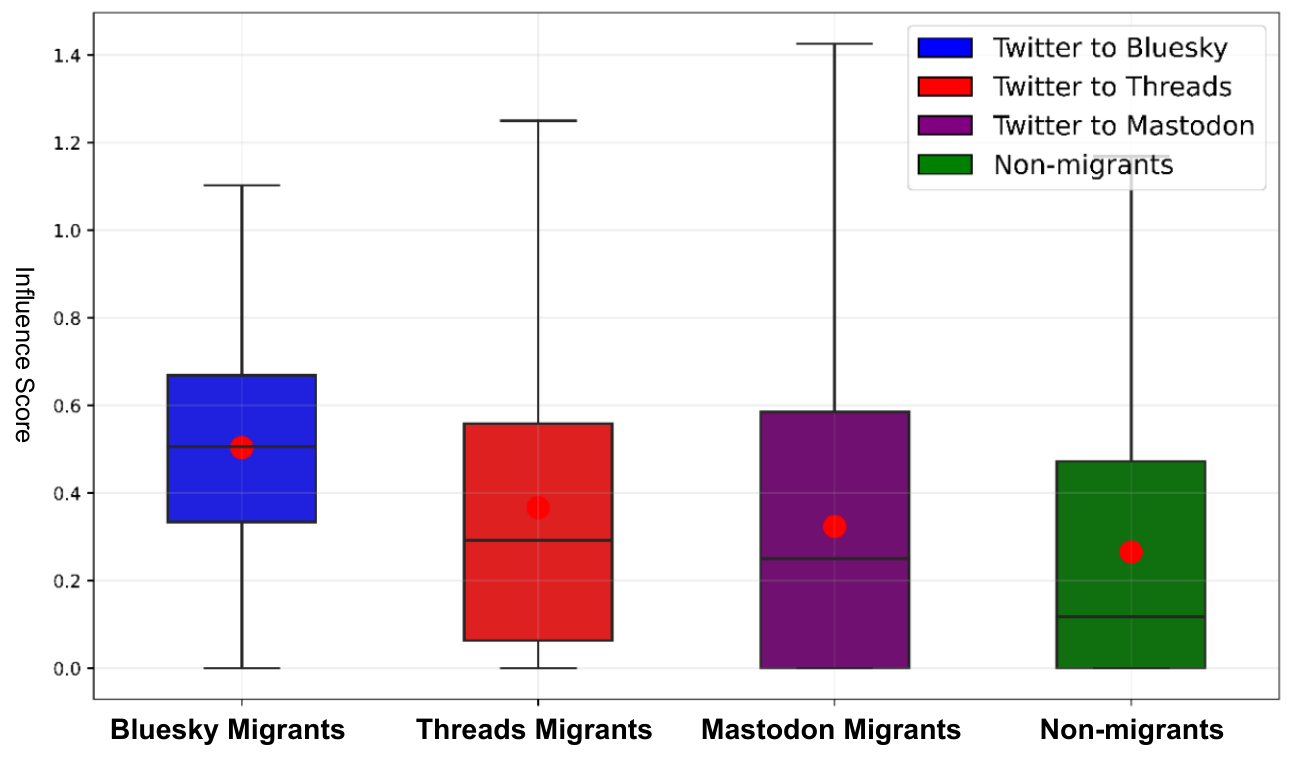}
\caption{\small Box plots display influence scores for migrant groups (Bluesky, Threads, Mastodon) and non-migrants (Twitter), highlighting their interquartile ranges. The red dots indicate the mean influence score for each group.}
\label{user_metric}
\end{figure}

\subsection{Comparing Influence Metrics.}


Building upon a prior methodology that assessed user influence on Twitter~\cite{anger2011measuring}, we calculate the Influence Score (IS) using two key metrics: the Interaction Ratio (IR) and the Retweets and Favorites Ratio (RFr). The IR is calculated by comparing the number of followers a user, denoted as $u$, has, with the number of times other users have retweeted or mentioned the given user. The RFr is determined by the proportion of the user's total tweets that have been either marked as retweeted or favorited.



\vspace{-0.25cm}

\begin{equation}
\begin{aligned}
& \text{Ir}(u) = \frac{\#retweets + \#mentions}{\#followers},\\
& \text{RFr}(u) = \frac{\#retweeted + \#favorited}{\#tweets},\\
&  \text{IS}(u) = \frac{\text{IR}(u) + \text{RFr}(u)}{2}
\end{aligned}
\end{equation}


Figure~\ref{user_metric} compares the influence scores of migrant groups on Bluesky, Threads, and Mastodon with those of non-migrants on Twitter, revealing three key insights. First, migrating users have higher mean influence scores, indicating that Twitter's more engaged users may consider other platforms. Second, distinct influence score patterns across these platforms suggest they cater to varied user preferences. Third, migrants to Bluesky, in particular, tend to retain higher engagement levels on Twitter than those migrating to other platforms.


\begin{tcolorbox}[colback=black!5!white,colframe=black!75!black,title=Summary (RQ1),rounded corners]
Migrant groups' varied characteristics show each platform attracted its distinct audience. Though migrants had a stronger Twitter presence than non-migrants, they also explored new platforms.
\end{tcolorbox}

\begin{figure*}
\centering
\includegraphics[width=1.0\textwidth]{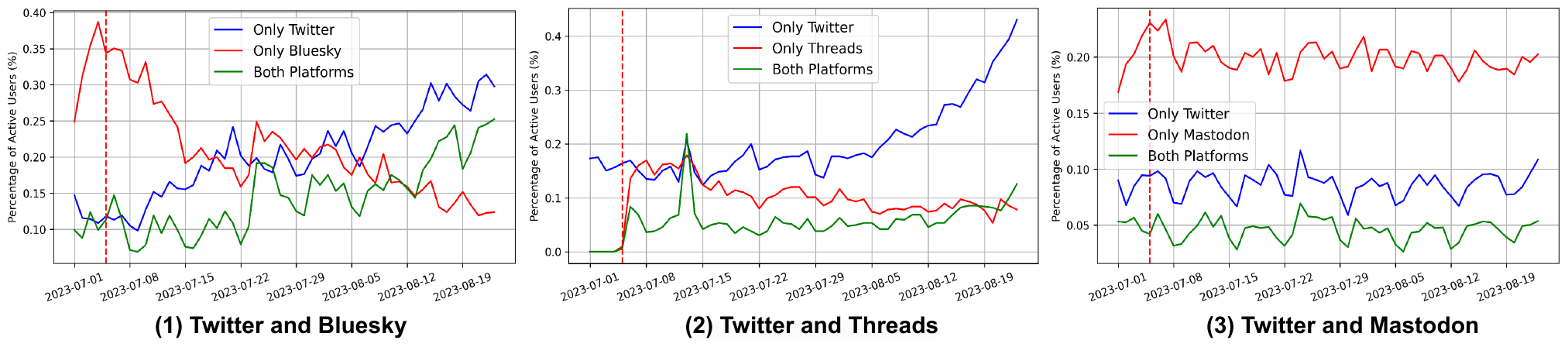}
\caption{\small Active user trends comparing Twitter with (1) Bluesky, (2) Threads, and (3) Mastodon. The blue line indicates users exclusively active on Twitter, the red line represents those only active on the alternative platform, and the green line denotes users active on both Twitter and the alternative platform. The red dashed line marks the launch date of Threads. The y-axis shows the percentage of active users relative to the total migrants in each category of active status on platforms.}
\label{active_trend}
\end{figure*}

\section{Understanding Relationships between Twitter and Alternative Platforms (RQ2)}

We conducted an analysis comparing individuals active on Twitter to those on alternative platforms. Our goal is to discern the dynamics and relationships between Twitter and its competitors as they vie for users' attention in the competitive landscape of social media.

\subsection{Comparing Active Users between Twitter and Alternatives.}
To understand how competition evolved between Twitter and its alternative platforms, we begin by counting the number of migrants who are active on each platform, using the  following definition:

\vspace{0.25cm}
\paragraph{\textbf{Active User}} For a platform \( p \), let a user be \( u \in U_p \). Given a time interval \( \delta = t' - t \) where \( t' > t \), the user \( u \) is active on platform \( p \) at time \( t' \) if the user engaged in posting or resharing since time \( t \).

\vspace{0.25cm}


Figure~\ref{active_trend} depicts the trend in the number of active users between Twitter and alternative platforms among the studied migrants. Initially, migrants from Twitter to Bluesky favored exclusive Bluesky usage, and this trend held strong until July 18, 2023, with a marked decrease afterward. Conversely, the number of users either staying dedicated to Twitter or using both platforms saw a consistent increase. This suggests that relying solely on Bluesky did not fully cater to users' needs.

Second, from the launch of Threads on July 5, 2023, until July 13, 2023, there is a consistent rise in the number of active users either adopting Threads exclusively or using it alongside Twitter. After just a week, the number of active users on Threads began to decrease, illustrating the ``shiny object effect'', where people are initially drawn to novelty, experience momentary joy in acquiring it, only to soon encounter difficulty in adopting the technology until they find meaningfulness~\cite{van2008newness}.

\vspace{0.5cm}

Last, the active users between Twitter and Mastodon do not show any dramatic changes for either platform. This stasis may be because users already experienced mass migration from Twitter to Mastodon, especially after Elon Musk's takeover of Twitter on October 27, 2022~\cite{jeong2023exploring}. The limited user overlap indicates that Mastodon operates independently of Twitter, and migrants to Mastodon typically divide into groups focused either on Twitter or Mastodon.

\subsection{Evaluating the Platform-level Association between Twitter and Alternatives.}
To understand the associations between Twitter and other platforms through the number of active migrants, we utilized Yule's Q, a statistical measure for assessing the association between two or more binary or nominal variables~\cite{yule1912methods}. We used this measure to analyze the presence or absence of users on Twitter compared to its competing platforms. Yule's Q offers valuable insights into whether Twitter usage reflects or influences behavior on the other platforms. On a scale ranging from -1 to 1, values close to 1 indicate a complementary relationship, and values nearing -1 suggest a substitute relationship.

To define Yule's Q in terms of a specific time period, we segmented the cumulative count of active users within interval \( \delta \) starting from date \( t \). We gauged active users on platforms $A$ and $B$ based on the subsequent metrics: \( U_{A, t, \delta} \) is the number of users who only used Platform $A$, not appearing on Platform $B$. Conversely, \( U_{B, t, \delta} \) is the number of users using only Platform $B$, absent on Platform $A$. Meanwhile, \( U_{AB, t, \delta} \) is the number of users using both platforms and \( U_{\neg A \neg B, t, \delta} \) is the number of users not using both platforms.

\begin{equation}
Q_{t, \delta} = \frac{{(U_{AB, t, \delta} \times U_{\neg A \neg B, t, \delta}) - (U_{A, t, \delta} \times U_{B, t, \delta})}}{{(U_{AB, t, \delta} \times P_{\neg A \neg B, t, \delta }) + (U_{A, t, \delta} \times U_{B, t, \delta})}}
\end{equation}


\begin{figure*}
\centering
\includegraphics[width=1.0\textwidth]{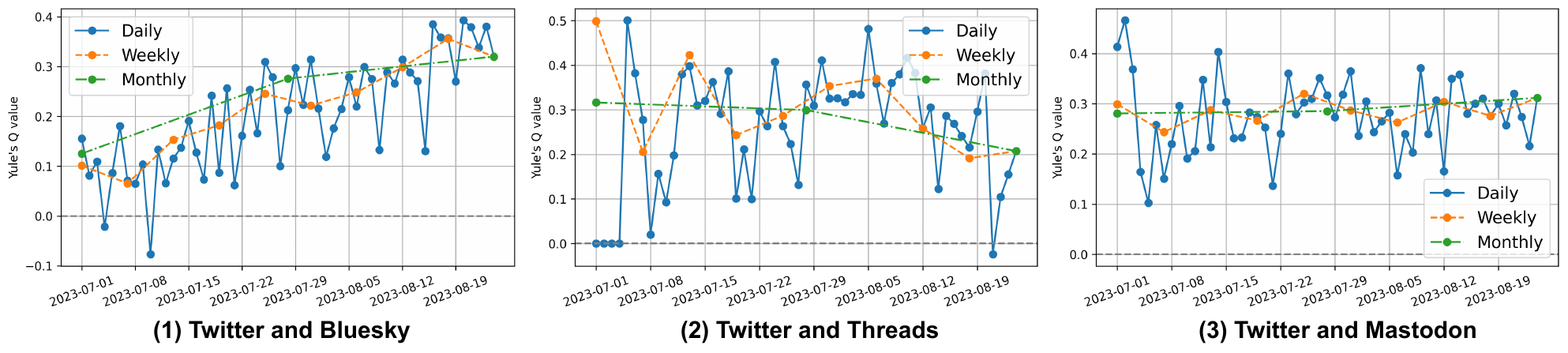}
\caption{\small Yule's Q trend between Twitter and (1) Bluesky, (2) Threads, and (3) Mastodon with three different intervals. Blue represents daily trends ($\delta= \text{1 day}$), orange for weekly trends ($\delta= \text{1 week}$), and green for monthly trends ($\delta= \text{4 weeks}$).}
\label{yule_Q}
\end{figure*}


Figure~\ref{yule_Q} depicts the evolution of Yule's Q over daily, weekly, and monthly intervals. The Yule's Q for Twitter and Bluesky consistently rises across intervals, transitioning from approximately 0.1 to around 0.3, highlighting Bluesky's complementary relationship with Twitter and its benefit from this association. Twitter and Threads display fluctuating Yule's Q values at both daily and weekly intervals. However, the monthly trend reveals a noticeable drop from approximately 0.3 to 0.2, suggesting a shift in user preference back towards Twitter. The relationship between Twitter and Mastodon remains steady, centering around Yule's Q value of 0.3 across all intervals, suggesting a complementary role and a stable migration pattern for Mastodon users.


\begin{tcolorbox}[colback=black!5!white,colframe=black!75!black,title=Summary (RQ2),rounded corners]

\small Migrants explored alternative platforms, but many could not show long-term success, leading migrants to return to Twitter. Yet, Bluesky migrants perceive Twitter as a primary complement, leading to distinct activity levels compared to Threads and Mastodon.
\end{tcolorbox}


\vspace{0.01cm}

\section{Did Migrants Really Leave Twitter? (RQ3)}

Frequent use of a platform does not necessarily signify user satisfaction, as many users continue to use the platform due to a lack of alternatives. In this section, we investigated the relationship between user activities and migrants' brand loyalty towards social media platforms.

\subsection{Brand Loyalty of Migrants To Platforms.}


Considering the noted uncertainty, we assessed the brand loyalty of users towards each platform, through textual analysis of their posts. Due to the lack of available datasets specifically annotated for brand loyalty of users, we leveraged ChatGPT (\texttt{gpt-4}) for classifying the stances into loyalty, neutrality, or disloyalty, given ChatGPT’s proven proficiency in stance detection~\cite{ziems2023can}. 

With our custom prompt designed for a target-based stance detection task (detailed in the Appendix), we assessed migrants' brand loyalty to the studied platforms. To this end, we extracted posts that mentioned specific platforms, grouped them by user, and concatenated these posts chronologically. To ensure viability, two coders annotated the stances of a randomly selected sample of 400 users, achieving a significant Cohen's Kappa coefficient of 76.27\%. Our method's effectiveness in this task was validated by F1 score of 79.42\%.

\begin{figure}[h]
\centering
\includegraphics[width=0.44\textwidth]{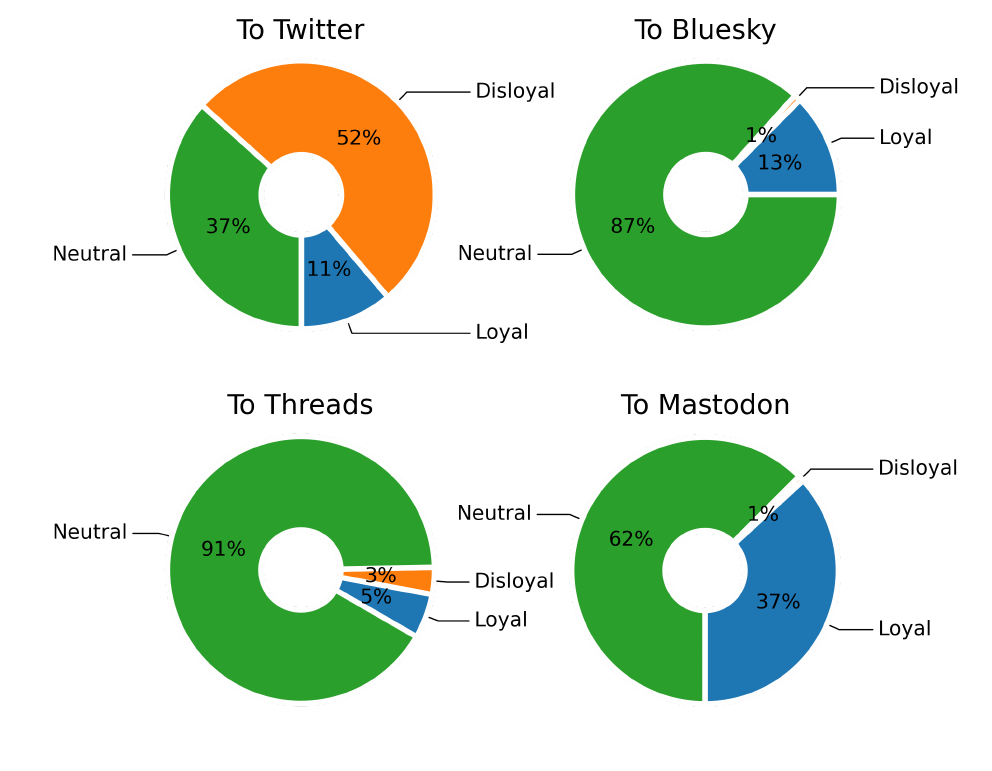}
\caption{\small The brand loyalty among migrants on various platforms, determined by stance from their aggregated posts.} 
\label{loyalty_piechart}
\end{figure}

Figure~\ref{loyalty_piechart} presents the distribution of brand loyalty among users across various platforms. Mastodon leads with 37\% loyalty among its migrants, surpassing figures from Twitter (11\%), Bluesky (13\%), and Threads (5\%). Notably, Twitter has the highest level of expressed disloyalty (52\%) among migrants. However, the majority of migrants on Bluesky, Threads, and Mastodon remain neutral, in contrast to a wider spectrum of stances on Twitter. This suggests that users are currently uncertain about their opinions on these newer platforms. 


\subsection{Rhetoric in Loyal and Disloyal Migrants.}

\begin{figure}
\centering
    \includegraphics[width=0.4838\textwidth]{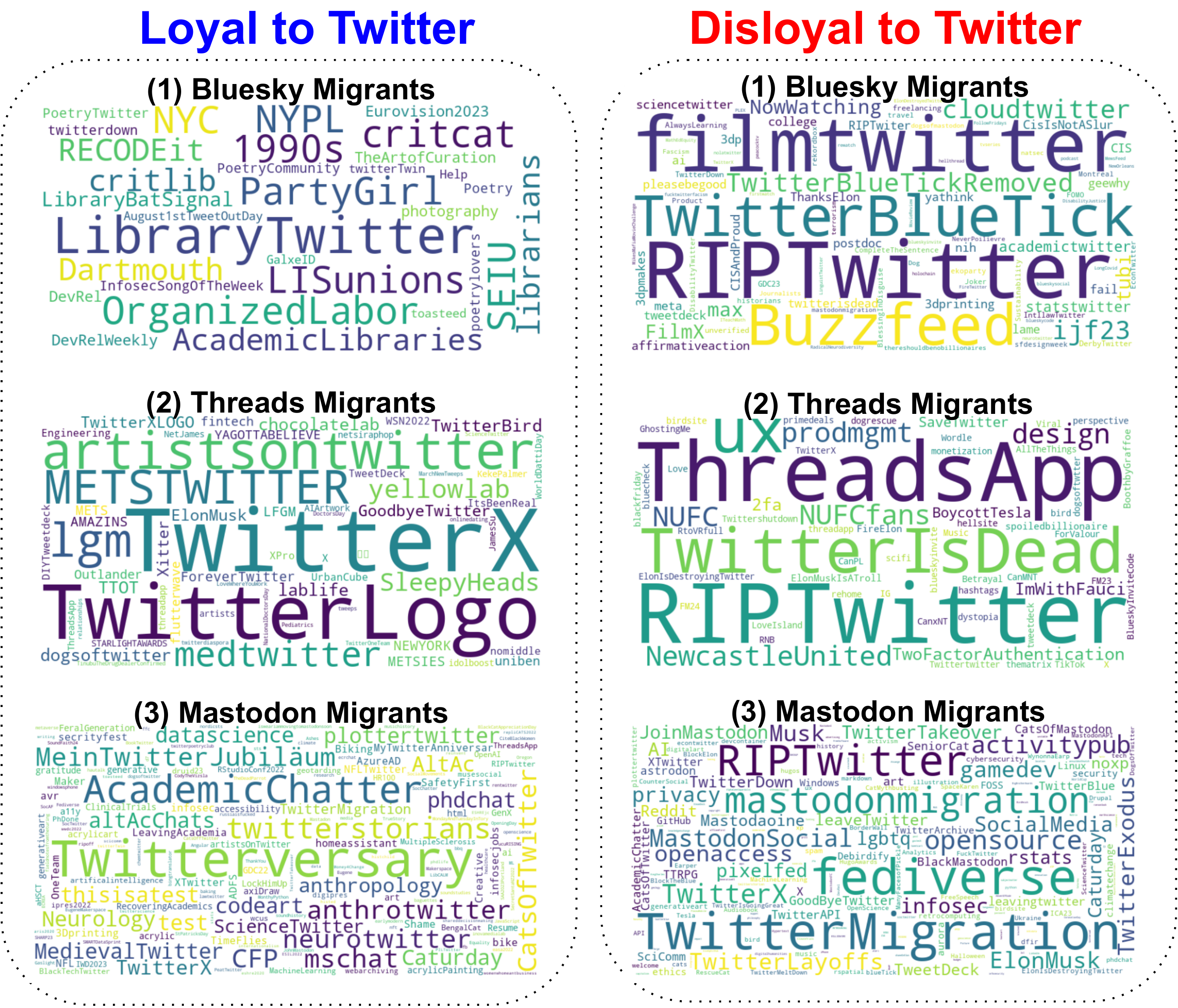}
\caption{\small Word clouds of migrants' hashtags based on two stances towards Twitter: Loyal (Blue) and Disloyal (Red).}

\label{loyalty_hashtags}
\end{figure}

Figure~\ref{loyalty_hashtags} depicts the distribution of hashtags among different platform migrant groups. Unsurprisingly, a universal trend is evident across all disloyal migrants: hashtags such as \texttt{\#RIPTwitter} and \texttt{\#TwitterIsDead} dominate the conversation. Loyal Bluesky migrants showcase their group affinity with \texttt{\#LibraryTwitter} and \texttt{\#FilmTwitter} on loyal and disloyal migrants, respectively. Loyal Threads migrants, in contrast, emphasize their artistic and sports-related affinities through hashtags such as \texttt{\#ArtistsOnTwitter} and \texttt{\#MetsTwitter}, and also comment on Twitter's new brand name and its logo with \texttt{\#TwitterX} and \texttt{\#TwitterLogo}. Loyal Mastodon migrants showcase a broader academic spectrum among loyal migrants compared to Bluesky and Threads. However, their disloyal counterparts predominantly focus on migration-specific hashtags including \texttt{\#TwitterMigration} and \texttt{\#Fediverse}, and IT-related hashtags such as \texttt{\#Opensource} and \texttt{\#ActivityPub}.




\subsection{Tendency in Platform Selection Among Disloyal Migrants.}

To measure the extent to which users struggle to leave Twitter despite an indication of disloyalty to Twitter, we conducted a Bayesian analysis centered on their activities across various platforms, including posting and resharing. Our analysis began by calculating the total number of activities undertaken by a user \( u \) on platform \( A \) at a given date \( t \), denoted as \( T_{A, u, t} \). To exclude cases where this count is zero, we introduce a smoothing factor by adding 1 to all $T_{A, u, t}$, resulting in an adjusted count represented by  $T'_{A, u, t}$.

We denote the prior probability of a user selecting platform $A$ as $\theta_A$. Since the selection of a platform is assumed to be binary for each activity, we model this prior probability with a Beta distribution, defined as:

\begin{equation}
\pi(\theta_A) = Beta(\alpha. \beta)
\end{equation}

\noindent where we choose both $\alpha$ and $\beta$ to be 1, as we have no initial preference or knowledge which results in a uniform distribution. By Bayes’ theorem, the posterior distribution of selecting platform $A$ can be written as:


\begin{equation}
P(\theta_A | T'_{A,u, t}) = \frac{P(T'_{A,u, t} | \theta_A) \pi(\theta_A)} {P(T'_{A,u, t})}
\end{equation}

To calculate the posterior probability, we adopt a Binomial distribution for the likelihood $P(T'_{A,u, t} | \theta_A)$, reflecting how often users choose platform $A$ versus their total activities on all platforms. Here, $n_t$, the number of trials, is the sum of activities on both platform $A$ and its counterpart at date $t$, shaping our likelihood as:

\begin{equation}
P(T'_{A,u, t} | \theta_A) = Binomial(n_t, \pi(\theta_A)) 
\end{equation}

\begin{figure*}
\centering
\includegraphics[width=0.975\textwidth]{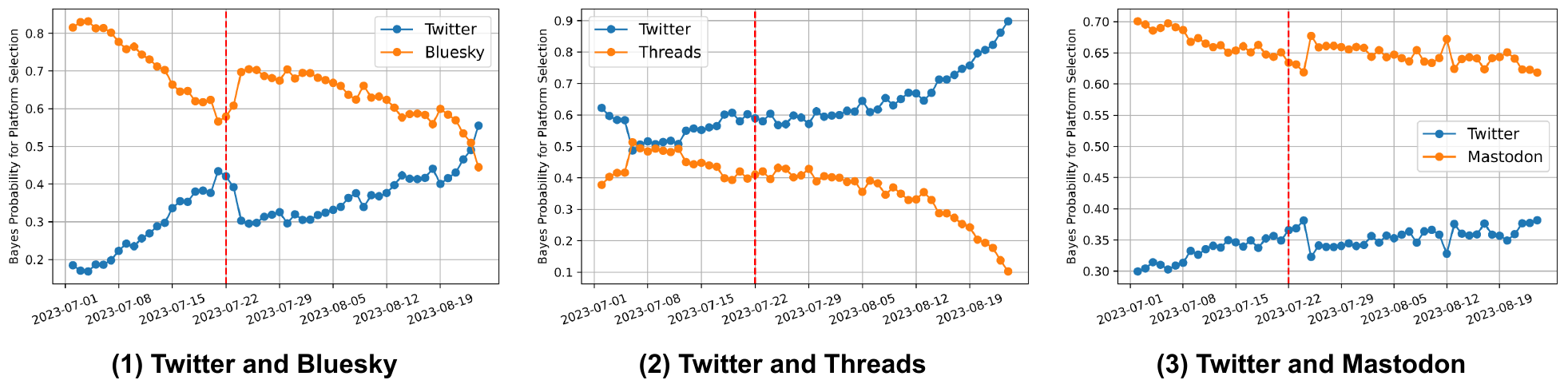}
\caption{\small Trends of platform selection among migrants disloyal to Twitter: We compared Twitter with its counterparts (1) Bluesky, (2) Threads, and (3) Mastodon. The blue line indicates the average posterior of individuals selecting Twitter over the counterpart, while the orange line is the opposite. The red dashed line marks when Twitter rebranded to ``$\mathbb{X}$''.} 
\label{disloyal_bayesian}
\end{figure*}

Figure~\ref{disloyal_bayesian} presents the average posterior probabilities of migrants showing disloyalty towards Twitter, highlighting distinct patterns among different platforms. First, there were two notable convergences in platform selection between Twitter and Bluesky, especially evident after Twitter's rebranding to ``$\mathbb{X}$" on July 22, with significant convergences occurring on this date and later on August 24. Secondly, a divergence was observed between Twitter and Threads, characterized by consistent returns of migrants to Twitter, with selection fluctuations starting on July 2, a minor dip on July 6, and an intensifying divergence after July 22. Last, a parallel trend existed between Twitter and Mastodon. Migrants to Mastodon consistently displayed a lower preference for Twitter, indicating a stabilization in their choice of platform, with a clear inclination towards Mastodon. This suggests that the migrants who have continued to stay on Mastodon until recently exhibit a reduced sensitivity to Twitter-related events, such as the rebranding, further emphasizing their commitment to Mastodon.

\begin{tcolorbox}[colback=black!5!white,colframe=black!75!black,title=Summary (RQ3),rounded corners]
 \small Migrants demonstrated a broader spectrum of brand loyalty towards Twitter than other platforms, often exhibiting signs of disloyalty towards Twitter. Despite this disloyalty, their activities on Twitter consistently overshadowed their engagement on all other platforms, Mastodon being the sole exception.
 \end{tcolorbox}
 

\section{Limitations}

First, our dataset is composed of migrants who voluntarily disclosed their identities on other platforms, suggesting a possible inclination towards more open communication and active engagement with peers. This potential selection bias could represent migrants with a more heightened online presence. Second, we centered on migration patterns from Twitter to other platforms, without considering migration between alternative platforms. This specific direction was chosen due to technical constraints, including the limited search features of Bluesky's API's and Threads's lack of a publicly accessible API, which hindered our ability to track migrations from Bluesky or Threads to other platforms. Last, we examined the first eight weeks after Threads' launch, which experienced a significant user influx\footnote{https://www.reuters.com/technology/metas-twitter-rival-threads-hits-100-mln-users-record-five-days-2023-07-10/}. This period might not capture earlier shifts from Twitter to Mastodon~\cite{jeong2023exploring}. Long-term analysis of Twitter data from these times was constrained by API pricing\footnote{https://developer.twitter.com/en/docs/twitter-api/getting-started/about-twitter-api}, imposing substantial fees for retrieving users' tweets and retweets.


\section{Conclusion \& Future Work}
Our analysis shows that Bluesky cleverly capitalized on the conflicts between Twitter and two of its competitors, such as Threads and Mastodon. With a lot of overlap and association in usage with Twitter's user base, Bluesky secured its spot in the competitive landscape. New platforms, such as Threads, initially benefit from the ``shiny object effect,'' attracting users with their newness. However, retaining these early enthusiasts proves challenging, especially when many still remain active on Twitter. While some initial migration barriers might be short-lived, the enduring pull of established platforms like Twitter is evident. Even if users voice intentions to switch, deep-seated inertia often keeps them anchored, either out of habit or due to perceived shortcomings in newer platforms.

In future work, we will explore attitudes towards Twitter and its proprietor. While a portion of its user base longs for the older version of Twitter, different segments of the user base already show highly varied levels of dissatisfaction with Twitter’s owner. We will also probe the structural determinants of brand loyalty on social platforms. Factors such as user interface design, social media fatigue, the prevalence of disinformation~\cite{jeong2022nothing}, and distinct social interactions unique to each platform~\cite{jeong2023exploring} can greatly sway user choices. To validate these findings, we will conduct online surveys using opt-in panels. Finally, we will delve deeper into the facets of brand loyalty, examining satisfaction, emotional connections, and perceived platform value, to understand the forces that keep users loyal or drive them away.




\section{Data Collection Policy}
We collected data from the designated social media platforms using their public interfaces. We are aware of the alteration to Twitter's terms of service\footnote{https://twitter.com/en/tos}, effective September 29, 2023. Following their guideline for using the currently published interfaces by Twitter, our data collection strictly employed Twitter's official API. Furthermore, we manually sourced text data from Threads due to the absence of its public API. We ensured user privacy by anonymizing personal data during our analysis. The code for mapping migrants between Twitter and other platforms (Bluesky, Threads, and Mastodon), as well as the data for user IDs of these migrants, is available at \url{https://github.com/ujeong1/SDM24_user_migration_across_multiple_platforms}.

\section{Acknowledgments}
This work received support from the Office of Naval Research, under Award No. N00014-21-1-4002. Opinions, interpretations, conclusions, and recommendations within this article are solely those of the authors.

\bibliographystyle{sdm24}

\bibliography{references}

\begin{thebibliography}{10}

\bibitem{anger2011measuring}
{\sc I.~Anger and C.~Kittl}, {\em Measuring influence on twitter}, in Knowledge
  Management and Knowledge Technologies, 2011.

\bibitem{auxier2021social}
{\sc B.~Auxier and M.~Anderson}, {\em Social media use in 2021}, Pew Research
  Center,  (2021).

\bibitem{azose2019estimation}
{\sc J.~J. Azose and A.~E. Raftery}, {\em Estimation of emigration, return
  migration, and transit migration between all pairs of countries}, PNAS,
  (2019).

\bibitem{cava2023drivers}
{\sc L.~L. Cava, L.~M. Aiello, and A.~Tagarelli}, {\em Drivers of social
  influence in the twitter migration to mastodon}, Scientific Reports,  (2023).

\bibitem{fiesler2020moving}
{\sc C.~Fiesler and B.~Dym}, {\em Moving across lands: Online platform
  migration in fandom communities}, Human-Computer Interaction,  (2020).

\bibitem{givon1984variety}
{\sc M.~Givon}, {\em Variety seeking through brand switching}, Marketing
  Science,  (1984).

\bibitem{he2023flocking}
{\sc J.~He, H.~B. Zia, I.~Castro, A.~Raman, N.~Sastry, and G.~Tyson}, {\em
  Flocking to mastodon: Tracking the great twitter migration}, in Internet
  Measurement Conference, ACM, 2023.

\bibitem{hou2011migrating}
{\sc A.~C. Hou, C.-C. Chern, H.-G. Chen, and Y.-C. Chen}, {\em ‘migrating to
  a new virtual world’: Exploring mmorpg switching through human migration
  theory}, Computers in Human Behavior,  (2011).

\bibitem{hou2020understanding}
{\sc A.~C. Hou and W.-L. Shiau}, {\em Understanding facebook to instagram
  migration: a push-pull migration model perspective}, Information Technology
  \& People,  (2020).

\bibitem{jeong2022nothing}
{\sc U.~Jeong, K.~Ding, L.~Cheng, R.~Guo, K.~Shu, and H.~Liu}, {\em Nothing
  stands alone: Relational fake news detection with hypergraph neural
  networks}, in IEEE Big Data, IEEE, 2022.

\bibitem{jeong2023exploring}
{\sc U.~Jeong, P.~Sheth, A.~Tahir, F.~Alatawi, H.~R. Bernard, and H.~Liu}, {\em
  Exploring platform migration patterns between twitter and mastodon: A user
  behavior study}, arXiv preprint arXiv:2305.09196,  (2023).

\bibitem{keller1993conceptualizing}
{\sc K.~L. Keller}, {\em Conceptualizing, measuring, and managing
  customer-based brand equity}, Journal of marketing,  (1993).

\bibitem{kumar2011understanding}
{\sc S.~Kumar, R.~Zafarani, and H.~Liu}, {\em Understanding user migration
  patterns in social media}, in AAAI, AAAI Press, 2011.

\bibitem{lee1966theory}
{\sc E.~S. Lee}, {\em A theory of migration}, Demography,  (1966).

\bibitem{levitt2007transnational}
{\sc P.~Levitt and B.~N. Jaworsky}, {\em Transnational migration studies: Past
  developments and future trends}, Annual Review of Sociology,  (2007).

\bibitem{matias2016going}
{\sc J.~N. Matias}, {\em Going dark: Social factors in collective action
  against platform operators in the reddit blackout}, in Human Factors in
  Computing Systems, 2016.

\bibitem{monti2023online}
{\sc C.~Monti, M.~Cinelli, C.~Valensise, W.~Quattrociocchi, and M.~Starnini},
  {\em Online conspiracy communities are more resilient to deplatforming},
  arXiv preprint arXiv:2303.12115,  (2023).

\bibitem{rogers2020deplatforming}
{\sc R.~Rogers}, {\em Deplatforming: Following extreme internet celebrities to
  telegram and alternative social media}, European Journal of Communication,
  (2020).

\bibitem{torkjazi2009hot}
{\sc M.~Torkjazi, R.~Rejaie, and W.~Willinger}, {\em Hot today, gone tomorrow:
  on the migration of myspace users}, in ACM workshop, 2009.

\bibitem{valero2023thousands}
{\sc M.~V. Valero}, {\em Thousands of scientists are cutting back on twitter,
  seeding angst and uncertainty}, Nature,  (2023).

\bibitem{van2008newness}
{\sc H.~C. van Trijp and E.~van Kleef}, {\em Newness, value and new product
  performance}, Trends in food science \& technology,  (2008).

\bibitem{yule1912methods}
{\sc G.~U. Yule}, {\em On the methods of measuring association between two
  attributes}, Journal of the Royal Statistical Society,  (1912).

\bibitem{zengyan2009cyber}
{\sc C.~Zengyan, Y.~Yinping, and J.~Lim}, {\em Cyber migration: An empirical
  investigation on factors that affect users' switch intentions in social
  networking sites}, in System Sciences, IEEE, 2009.

\bibitem{ziems2023can}
{\sc C.~Ziems, W.~Held, O.~Shaikh, J.~Chen, Z.~Zhang, and D.~Yang}, {\em Can
  large language models transform computational social science?}, arXiv
  preprint arXiv:2305.03514,  (2023).

\end{thebibliography}

\begin{center}
\noindent\textbf{Appendix}
\end{center}

Utilizing OpenAI API's function calling feature, we designed a prompt for a structured input and output in a few-shot setting, providing loyal and disloyal examples. This incorporated information about Twitter's rebranding to ``$\mathbb{X}$'', and the launch of Meta's new social media named Threads. We also confirmed that GPT-4 is already aware of Bluesky (a decentralized social media co-founded by Jack Dorsey) and Mastodon (a federated microblogging service founded by Eugen Rochko).
 %

 %
 %
\begin{promptbox}
\footnotesize
\textbf{Title: Stance Detection Prompt for GPT-4}

\textbf{Objective:}\\
Determine the stance of a given text towards a specified target platform.

\textbf{Instructions: }\\
For the provided text and target(s), classify the stance as one of the following: Loyal, Disloyal, or Neutral

\textbf{Keynotes:}\\
- There are four platforms that can be targeted for the stance: Twitter, Bluesky, Threads, and Mastodon.\\
- Twitter is now called ``$\mathbb{X}$'' and is owned by Elon Musk.\\
- ``Threads'' is a new social media platform under the Meta umbrella, founded by Mark Zuckerberg.\\
- Always provide a stance for each specified target.\\
- Provide the response in JSON format.

\textbf{Example:}\\
Input: ``Twitter is dead. I love Bluesky''\\
Output: \{Twitter: Disloyal, Bluesky: Loyal\}
\end{promptbox}

\end{document}